\documentclass[epj]{svjour}

\usepackage{latexsym}
\usepackage{amsmath}
\usepackage{amssymb}
\usepackage{graphicx}
\usepackage{graphics}
\usepackage{bm}
\usepackage{enumerate}

\begin{document}
\title{
Coupled-channel description for mirror mass-11 nuclei
compared to shell-model structures
}

\author{K. Amos
\inst{1,2}
\and
S. Karataglidis
\inst{2,1}
\and
L. Canton
\inst{3}
\and
P. R. Fraser
\inst{4}
\and
K. Murulane
\inst{2}
}

\institute{
School  of Physics,  University of  Melbourne,
    Victoria 3010, Australia.
    \email{amos@unimelb.edu.au}
\and
Department of Physics, University of Johannesburg,
    P.O. Box 524, Auckland Park, 2006, South Africa.
\and
Istituto  Nazionale  di  Fisica  Nucleare,
    Sezione di Padova, Padova I-35131, Italia.
\and
School of Science/Learning and Teaching Group,
The University of New South Wales, Canberra, ACT 2600, Australia.
}

\date{today}

\abstract{
The spectra of mass-11 nuclei are unusual, and so pose a challenge for 
nuclear-structure theory.  Relating to nucleon emission,
the set of isobars range from being well-bound
($^{11}$B,$^{11}$C) through weakly bound ($^{11}$Li, $^{11}$Be), 
to being proton unstable ($^{11}$N,$^{11}$O). 
To add complexity, the weakly bound $^{11}$Li takes the form of
a two-nucleon halo exotic.  A self-consistent
approach to understand this set of nuclei is especially important as 
the mirror pair $^{11}$Be-$^{11}$N exhibit a parity-inverted
ground state compared to their neighboring nuclei. 
Herein,
the Multi-Channel Algebraic Scattering  method (MCAS) has
been used to describe the low excitation spectra of those
isobars in terms of nucleon-nucleus clusters.
A collective model description of the low-excitation states of the mass-10
mass-10 core nuclei has been used to form the coupled-channel
interactions required in the method. For comparison, and to 
understand the underlying configurations, a shell model
approach has been used to obtain those spectra with no-core
$(0+2+4)\hbar\omega$ and $(0+2)\hbar\omega$ shell-model spaces for 
the mass 10 and mass 11 nuclei respectively.
The results of the calculations suggest the need of a strong
coupling in the collective coupled-channel vibrational model. 
In particular, the strong coupling of the collective $2^+_1$ state 
of $^{10}$Be to the valence neutron plays a decisive role in forming 
the positive parity ground state in 
$^{11}$Be; an effect confirmed by the shell-model results.
}


\maketitle

 
\section{Introduction}

The set of mass-11 isobars, from $^{11}$Li to $^{11}$O and encompassing three 
mirror pairs, span the region from the proton to
the neutron drip lines, with $^{11}$N and $^{11}$O being beyond the proton 
drip line. A systematic study of this system
is of interest given the change in isospin from one end of the range to the 
other, but also given that the mirror pair
$^{11}$Be and $^{11}$N have ground states of positive parity, opposite to 
those of the other nuclei in the system. That has
been the subject of extensive studies (see, for example, 
Refs. \cite{Sa93,Ca16,Ba17}), as $^{11}$Be is also 
considered to be a single-neutron halo.
While the determination of the nature of the positive-parity ground states 
is of interest, it is important to
understand those ground states also with regards to the neighbouring mass-11 
nuclei. In that context, a self-consistent many-body
interpretation across the mass-11 isobars is necessary. In so doing, it 
becomes necessary to understand the nature of
the ground state of $^{11}$Be beyond the extreme single-particle picture as 
is suggested by the halo,
which would require that its ground state wave function has significant 
neutron occupancy in the $1s_{1/2}$ orbit. 

These mass-11 nuclei have quite diverse nucleon emission thresholds as is 
shown in Table~\ref{Tabthr}. 
They range from 22.39 to $-$1.49~MeV. 
Recently \cite{We19}, an observation of the unbound $^{11}$O 
was made and of its decay
by two proton emission.
Describing the spectra of such a set of odd-mass nuclei self-consistently 
presents a challenge.
Herein we compare spectra found using a
coupled-channel description for the mirror mass-11 nuclei
with results from complete-space shell-model calculations. We have used 
the coupled-channel approach \cite{Am03} that, previously, was
taken to similarly analyse  the spectra of the full 
set of mass-7 isobars \cite{Ca06}.

Those analyses assumed two types of coupling: ones involving two clusters 
of composite nuclei (mass-3 and mass-4) for 
only a single channel, and the other, coupling of nucleons to low-lying states 
in $^6$He and $^6$Be. For the former,
the assumption of only a single channel contributing to the spectrum is 
adequate given that there are no excited states in the 
core nuclei, save for the resonances above 20 MeV in $^{4}$He \cite{Ti92}, 
which do not affect the low-lying spectra of 
the compound.
\begin{table}
\caption{\label{Tabthr} General properties of mass-11 isobars \cite{Ke12}.
Uncertain/unknown values shown in brackets. Energies are in MeV.}
	\scalebox{1.0}{%
\begin{tabular}{c|cccccc}
\hline
\rule[-0.2cm]{0pt}{0pt}  \rule{0pt}{0.4cm}
 & $^{11}$Li & $^{11}$Be & $^{11}$B & $^{11}$C & $^{11}$N $(^{\ast})$ & 
$^{11}$O\\
\hline
\rule{0pt}{0.4cm} 
$J^\pi$ (gs) 
& $\frac{3}{2}^-$ & $\frac{1}{2}^+$ & $\frac{3}{2}^-$ & $\frac{3}{2}^-$ 
& $\frac{1}{2}^+$ & ($\frac{3}{2}^-$) \\
 decay & $\beta^-$ & $\beta^-$ & stable & $\beta^+$ &  
$1p$ & $2p$ \\
$1p$ thr. & 15.73 & 20.16 & 11.23 & 8.69 &  $-1.49$ & 
(---) \\
\rule[-0.2cm]{0pt}{0pt}
$1n$ thr. &  0.396 & 0.502 & 11.45 & 13.12 & 22.39 & (---) \\
\hline
\end{tabular}}
.\\
$^\ast$ In~\cite{Ca06} a proton separation threshold of $-$1.54~MeV
was suggested for $^{11}$N.
\end{table}

Our interest is to describe the low excitation spectra of 
the set of  mirror pairs, 
$^{11}$C-$^{11}$B and 
$^{11}$Be-$^{11}$N, within a coupled-channel approach. 
Too little is known about the third mirror pair, $^{11}$Li-$^{11}$O, and 
their underlying mass-10
cores for the method to be used with any confidence.
We seek to establish the extent to which a nearly-uniform interaction may 
account for the spectra of 
four of these isobars. For these cases, we treat the mirror pairs as the 
coupling 
of a nucleon with the appropriate 
mass-10 core, which is described by a collective model (either vibrational 
or rotational).

The method we use solves the coupled-channel problems of two-cluster systems, 
usually nucleon-nucleus, algebraically.
The coupling interactions in this approach have been formed using a number of 
low 
excitation states of the core nuclei.
In those coupled-channel calculations, the basic nucleon-nucleus interaction 
used has the form
\begin{equation}
v_0(r) = 
\left[ V_0 + V_{ll} \left\{ \mathbf {l \cdot l} \right\} \right]
w(r) + 2\lambda_\pi^2 
V_{ls} \frac{1}{r} \frac{\partial w(r)}{\partial r} 
\left\{ \mathbf{l} \cdot \mathbf{s} \right\}\; ,
\label{poteq}
\end{equation}
in which the Woods-Saxon (WS) function,
\begin{equation*}
w(r) = \left[1 + \exp\left(\frac{r - R_0}{a_0} \right) \right]^{-1}, 
\end{equation*}
has been used. 
The vector operators $\textbf{l}$ and $\textbf{s}$ denote orbital and 
nucleon intrinsic spin, and $V_0$, $V_{ll}$, $V_{ls}$ are potential-strength 
parameters. $R_0$ is nuclear radius and $a_0$ is the diffusivity.
All details of the approach, first used in 
\cite{Am03}, and a collection of the results of applications
made to date are given in ref.~\cite{Ka19}.

The Pauli principle is taken into account in this method by 
adding orthogonalizing pseudo-potentials (OPP) to the 
nucleon-nucleus potentials chosen for the cluster 
coupled-channels interactions in the defining Hamiltonian 
for the cluster \cite{Ca05}.  The OPP are formed as nonlocal separable
products of bound single nucleon wave functions in the 
chosen nucleon-nucleus potentials.  Thus the interaction
matrix of potentials used has the non-local form,
\begin{equation}  
{\cal V}_{cc'}(r,r') = V_{cc'}(r)\ \delta(r-r') + \lambda A_{c}(r)
A_{c}(r')\ \delta_{c,c'}\ ,
\end{equation}
in which $V_{cc'}$ are the collective-model derived nuclear potentials, 
$A_c(r)$ is the radial part of the single particle wave function in 
channel $c$, determined by solving the radial Schr\"odinger equation, 
and $\lambda$ is a scaling energy. 
A sufficiently large value for these energy parameters (about $10^6$ MeV)
implies that completely-occupied nucleon sub-shells in the target
become inaccessible to occupancy by a projectile nucleon.
Smaller non-zero values lead to a Pauli-hindrance effect involving a
partially occupied sub-shell, as has been extensively discussed 
~\cite{Ka19,Ca06x,La19}.
Pauli-allowed states imply the scaling parameter being set to zero.

For completeness,
we have also evaluated the spectra of 
some of the mass-10 and mass-11 
isobars of relevance using a complete  space, no-core, shell model. 
For mass-10 
we have used a $(0+2+4)\hbar\omega$ shell model while for mass-11
a no-core $(0+2)\hbar\omega$ shell model was used. Use of the smaller
basis model for the odd-mass nuclei was due to the much larger matrices 
involved
in using the $(0+2+4)\hbar\omega$ model for mass-11 nuclei rendering
those calculations impractical. In all cases, the OXBASH shell-model 
program \cite{Ox86} was used. The results are free of any
spurious center of mass effects since the bases are complete.

Results are presented in two sections.  In
Sect.~\ref{sect2} we consider the structure of three sets 
of mass-11 mirror pairs as found using MCAS to solve 
coupled-channel cluster systems. Results of using this 
approach are given in three subsections. 
In Sect.~\ref{sect3}, spectra of the mass 10 
and 11 isobars that we have found using no-core shell 
models are presented and discussed.
Conclusions are given in Sect. \ref{conc}.

\section{Mass-11 isobars as cluster systems}
\label{sect2}

The MCAS method has been used to solve couple-channel equations to give
low-energy spectra of a set of mass-11 isobars. 
These systems are described by the effective 
coupling of a nucleon to a $^{10}$Be, a $^{10}$B, or a $^{10}$C core. The 
interactions
for the channel couplings were defined using collective models for the core 
nuclei.
The spectra for the mirror pair $^{10}$Be and $^{10}$C exhibit structures 
somewhat
consistent with a vibrational model (being a $0^+$ ground state, followed 
by almost-equal spacings to a $2^+$ excited state and then a near-degenerate 
triplet). However, given the lack of knowledge of the
spectrum of $^{10}$C, we restrict the couplings to involve the first
four states in each only. The low-lying spectrum of $^{10}$B does not show 
features consistent
with a vibrational model. We use the lowest six states of $^{10}$B
to define the nucleon-nucleus channel couplings in a rotational model.

Results from those calculations are presented and discussed in the following
three subsections for the mirror systems $^{11}$C (described as the coupling
of a neutron to $^{10}$C) and $^{11}$B $(p +\,^{10}\text{Be})$; $^{11}$N
$(p+\,^{10}\text{C})$ and $^{11}$Be $(n+\,^{10}\text{Be})$; and,
$^{11}$C $(p+\,^{10}\text{B})$ and $^{11}$B $(n+\,^{10}\text{B})$, sequentially.

\subsection{The mirror systems, $^{11}$C and $^{11}$B}

Initially, we consider $^{11}$C and $^{11}$B as modelled by
couplings of the clusters $n+^{10}$C and $p+^{10}$Be, respectively. The
spectra of these mirror systems have many bound states given the 
relevant emission thresholds as shown in Table~\ref{Tabthr}. 
In our case, using the coupled-channel approach, good results have been found 
assuming a vibration collective model and
for the relevant nucleon interactions, with the low-lying 
states of $^{10}$C and $^{10}$Be.  We have used the four 
known positive states to $\sim 6$~MeV excitation in each 
core nucleus. For the vibrational model, the assumed target states
are: the ground states ($0^+$), which are taken as the vacuum, the 
first excited states ($2^+$), assumed to be one 
quadrupole phonon excitation on the vacuum, while the other excited 
states ($2^+_2, 0^+_2$) are taken as two quadrupole 
excitations.

The parameter values defining the interactions are listed 
in Table~\ref{Param1}. Therein, all energies are in MeV and
the radii and diffusivities are in fm.
\begin{table}{%
\caption{\label{Param1} The parameter values
	of Eqs.~(1), (2) and (3) that have been  
	used in our coupled-channel
evaluations of the $^{11}$C (as $n+^{10}$C) and $^{11}$B (as $p+^{10}$Be) 
	spectra. Strengths are
in MeV, geometry radii and diffusivities are in fm.}
\begin{tabular}{cccc}
\hline
	\rule[-0.2cm]{0pt}{0pt} \rule{0pt}{0.4cm}
\rule{0pt}{0.4cm}
	Nuclear (WS)  &  $R_{0} = 2.8$   & $a_{0} = 0.65$ & \\
	& $V_0 = -53.3$ &  $V_{ls}= \phantom{-}6.2$ & $V_{ll} = 0.0$\\
	& & & $\beta_2 = -$0.83\\
\hline
\rule{0pt}{0.4cm}
\rule[-0.2cm]{0pt}{0pt}
Coulomb (3pF)  & $R_{c} = 2.39$ & $a_{c}=0.8$ & $w_c=-0.06$ \\
\hline
\end{tabular}}
\end{table}
It should be noted that in Table~\ref{Param1} the geometric Coulomb parameters 
are constrained
by analyses of the charge densities of the cores \cite{Bh00} as given 
by a three-parameter Fermi
distribution function {3pF} \cite{Ka19}, namely 
\begin{equation}
\rho_{\text{ch}}(r) =
\frac{1 + w_c \left( r^2/R^2_c \right)}{1 + \exp \left[ 
	\left( r - R_c \right)/a_c \right] } \; .
\end{equation}
In calculations, the parameters, radius, $R_c$,
diffuseness $a_c$, and scaling parameter $w_c$, 
are chosen to give the rms Coulomb charge radius listed
in Ref.~\cite{Bh00}.

The OPP, $\lambda^{(I)}_{nlj}$, required for those interactions with
states in the mass-10 core nuclei are listed in Table~\ref{OPPs}.
\begin{table}
\caption{\label{OPPs}
OPP strengths for relevant nucleon interactions with states
in $^{10}$C ($^{10}$Be). All energies \cite{Ke12} are in MeV
and those for the mass-10 states are the experimental ones.}
	\scalebox{0.9}{%
\begin{tabular}{cccccc}
	\hline
\rule[-0.2cm]{0pt}{0pt} \rule{0pt}{0.4cm}
state & $E_x$, $^{10}$C ($^{10}$Be) & Widths $^{10}$C &
$0s_{1/2}$ & $0p_{3/2}$ & $0p_{1/2}$ \\
\hline
\rule{0pt}{0.4cm}
g.s.    & 0.000 (0.000) & --- & $10^6$ & 5.2  & 9.0 \\
$2_1^+$ & 3.354 (3.368) & --- & $10^6$ & 10.0 & 9.0  \\
$2_2^+$ & 5.220 (5.958) & 0.225 & $10^6$ & 0.0  & 0.0  \\
\rule[-0.2cm]{0pt}{0pt}
$0_2^+$ & 5.380 (6.180) & 0.3 & $10^6$ & 0.0  & 0.0 \\
\hline
\end{tabular}}
\end{table}

\begin{figure}
\scalebox{0.35}{\includegraphics*{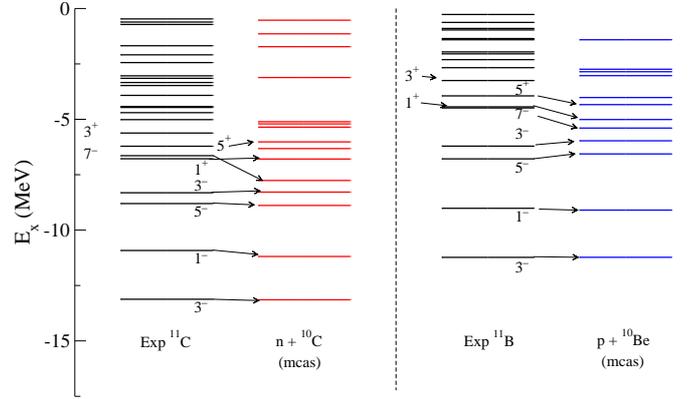}}
\caption{
\label{C11-B11}
Experimental spectra of $^{11}$C and of $^{11}$B~\cite{Ke12} 
compared with the results of MCAS calculations for the 
$n+^{10}$C and $p+^{10}$Be clusters.  The zero energy in each 
is the relevant nucleon separation energy.  The spin-parities  
of the states are indicated as $2J^\pi$.}
\end{figure}

In Fig.~\ref{C11-B11}, the results found for the low excitation spectra 
of $^{11}$C
and $^{11}$B are compared with the known spectra~\cite{Ke12}. 
Good agreement is evident over almost 10~MeV in excitation.
As good an agreement has been found with the same 
Hamiltonian plus a Coulomb interaction, when used in 
calculating the $^{11}$B spectrum. 

Our calculated results reproduce 
the observed nucleon separation energies, find the proper  
sequence of low excitation negative parity states 
(energy gaps and spins) and give  
the known excitation energies of the first positive parity 
excited states.

\subsection{The mirror systems, $^{11}$Be and $^{11}$N}

We have described $^{11}$N and $^{11}$Be by the coupling of the
clusters $p+^{10}$C and
$n+^{10}$Be, respectively. They pose very different
problems for evaluation to those of 
the $^{11}$C and $^{11}$B mirrors. 
Largely those different problems relate to
$^{11}$Be being weakly  bound while $^{11}$N  is unbound as
is indicated by the
relevant nucleon  thresholds listed in Table~\ref{Tabthr}.  
Further, both 
ground states (a resonance in the case of $^{11}$N) have assigned 
spin-parities of $\frac{1}{2}^+$.  

Our coupled-channel calculations for the spectra 
of $^{11}$N and $^{11}$Be
used the parameter values for the interactions that are listed in 
Table~\ref{Param2}, with a
deformation parameter $\beta_2 = -0.83$.
\begin{table}
\caption{\label{Param2} 
The parameter values used in the coupled-channel
evaluations of the $^{11}$N (as $p+^{10}$C) and $^{11}$Be (as $n+^{10}$Be) 
	spectra.
Strengths are in MeV, radii and diffusivities are in Fermi.}
\begin{tabular}{cccc}
\hline
\rule{0pt}{0.4cm}
\rule[-0.2cm]{0pt}{0pt}
Nuclear (WS) &  $R_0 =$ 2.8  & $a_0 =$ 0.625 & \\
	& $V_0 =-$42.2 & $V_{ls} = $6.0 & $V_{ll} = 0.7$\\
 & & & $\beta_2 = -$0.83\\
\hline
	\rule{0pt}{0.4cm}
\rule[-0.2cm]{0pt}{0pt}
	Coulomb (3pF)  & $R_c = $ 2.4 & $a_c = $0.8 & $w_c = $0.0 \\
\hline
\end{tabular}
\end{table}
While most of these parameter values are similar to those given
in  Table~\ref{Param1} for the $^{11}$C and $^{11}$B cases,
the values of the central strengths 
are $\sim 30\%$ smaller.  This change in the central potential strength reflects the
 changed number of like nucleon versus unlike nucleon interactions
 between the nucleons in each mass-10 core with the added nucleon
 forming the clusters.
 
The OPP strengths needed in these cases also differ from the values for the
clusters due to the changed numbers of like nucleons (to the external 
one) in the core nuclei. 
In particular, in the current configurations, the two ground states 
have the $0p_\frac{3}{2}$ orbits fully occupied. 
The OPP strengths required to 
obtain the low excitation spectra of $^{11}$N and $^{11}$Be 
are listed in Table~\ref{OPP2}; with that of the $0p_{\frac{1}{2}}$
orbit in the ground state of $^{10}$Be being slightly larger
to give the observed excitation energies of the  $\frac{1}{2}^-|_1$ states. 

\begin{table}
\caption{\label{OPP2}
OPP strengths for relevant nucleon interactions with states
in $^{10}$C ($^{10}$Be) to get the results shown in
Fig.~\ref{N11-Be11}.  All energies are in MeV and
	core nuclei excitation energies are the experimental
	values.}
	\scalebox{0.9}{%
\begin{tabular}{cccccc}
	\hline
\rule[-0.2cm]{0pt}{0pt} \rule{0pt}{0.4cm}
state & $E_x$,  $^{10}$C ($^{10}$Be) & Widths $^{10}$C &
$0s_{1/2}$ & $0p_{3/2}$ & $0p_{1/2}$ \\
\hline
\rule{0pt}{0.4cm}
	g.s.    & 0.000 (0.000) & --- & $10^6$ & $10^6$  & 5.3 (5.0) \\
$2_1^+$ & 3.354 (3.368) & --- & $10^6$ & 83.0 & 13.7  \\
$2_2^+$ & 5.220 (5.958) & 0.225 & $10^6$ & 4.4  & 3.0 (2.8)  \\
\rule[-0.2cm]{0pt}{0pt}
$0_2^+$ & 5.380 (6.180) & 0.3 & $10^6$ & 4.3  & 0.0 \\
\hline
\end{tabular}}
\end{table}

For optimal results in these two cases,
slightly differing central
potentials of strength of $-42.2$ and $-41.8$~MeV 
and diffusivities of 0.625 and 0.7~fm.
have been used to get the spectra of $^{11}$Be and $^{11}$N respectively.
The need of these small adjustments for mirror conditions, in this case, 
is not surprising given
the small differences in spectral properties of the two
mass-10 core nuclei as well as
the simplistic collective model used to describe the coupling
interactions between the nucleon and
the four states of those core nuclei chosen for the
coupled-channel calculations. 

\begin{figure}
\centering\scalebox{0.55}{\includegraphics*{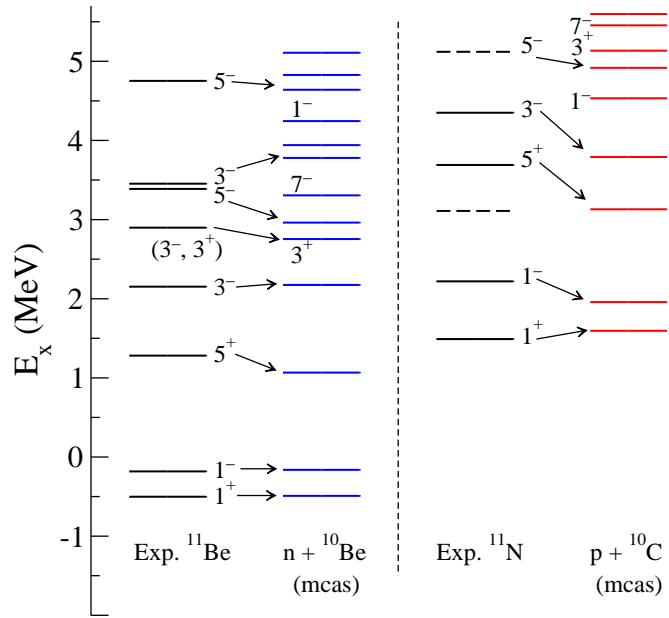}}
\caption{
\label{N11-Be11}
Experimental spectra of $^{11}$Be and of $^{11}$N \cite{Ke12}
compared with the results of MCAS calculations for
the $n+^{10}$Be and $p+^{10}$C clusters.
The zero energy in each is the relevant nucleon
separation energy.
The spin-parities of the states are indicated
as $2J^\pi$.}
\end{figure}

As shown in Fig.~\ref{N11-Be11},  our coupled-channel 
calculations 
gave the nucleon separation energies and the approximate energies 
and spin-parities of many of the known spectral states   
in the low  excitation spectra. But there are calculated levels
that are not observed and known levels that are not matched
in this (and higher) excitation energy range. 
Possibly there is a state at 3.06~MeV in the spectrum
of $^{11}$N \cite{Ke12} that is not found in the calculation.
But there has not been any spin-parity assigned 
to it in the tabulations. 
Properties of the spectra are listed in Table~\ref{Etable}.
Therein the experimental energies and widths of known states in  
the low excitation spectra of $^{11}$Be and $^{11}$N are compared.
\begin{table}
\caption{\label{Etable}
The known and MCAS results for $^{11}$Be and $^{11}$N
as given in the Fig.~\ref{N11-Be11}.
The experimental and MCAS energies and widths are compared.
	Level energies are in MeV, widths (in brackets) are in keV.}
	\scalebox{0.84}{%
	\begin{tabular}{ccc|ccc}
\hline
\rule[-0.2cm]{0pt}{0pt} \rule{0pt}{0.4cm}
	State & \multicolumn{2}{c}{$^{11}$Be} & 
	State & \multicolumn{2}{c}{$^{11}$N} \\
$J^\pi$ & Experiment & MCAS & $J^\pi$ & Experiment & MCAS \\
\hline
\rule{0pt}{0.4cm}
$\frac{1}{2}^+$ & $-$0.501 & $-$0.501 & $\frac{1}{2}^+$ & 1.49 (830) & 1.49 (2760)\\
$\frac{1}{2}^-$ & $-$0.18 & $-$0.182 & $\frac{1}{2}^-$ & 2.22 (600) & 2.21 (10)\\
		& & & & $3.06 (<100)$ & \\
$\frac{5}{2}^+$ & 1.28 (100) & 1.05 (118) & $\frac{5}{2}^+$ & 3.69 (540) & 3.19 (573)\\
$\frac{3}{2}^-$ & 2.15 (206) & 2.15 (96) & $\frac{3}{2}^-$ & 4.36 (340) & 4.08 (172)\\
$\frac{1}{2}^-$ & --- & 2.74 (11) &  $\frac{1}{2}^-$ & --- & 4.44 (3) \\
$(\frac{3}{2}^{(\pm)})$ & 2.95 (122) & 2.95 (275) & $\frac{3}{2}^+$ & --- & 4.87 (3200)\\
$\frac{5}{2}^-$ & 3.39 ($<8$) & 3.29 ($<1$) & $\frac{5}{2}^-$ & 5.12 ($<220$) & 5.09 (165)\\
$\frac{7}{2}^-$ & --- & 3.76 (12) & $\frac{7}{2}^-$ & --- & 5.57 (340) \\
$\frac{3}{2}^-$ & 3.46 (10) & 3.92 (131) & $\frac{5}{2}^-$ & 5.91 (?) & 6.41 (150)\\
$\frac{1}{2}^-$ & --- & 4.22 (107) & ($\frac{3}{2}^-)$ & 6.57 (100) & 6.51 (258) \\
\rule[-0.2cm]{0pt}{0pt}
$\frac{5}{2}^-$ & 4.76 (45) & 4.62 (88) & & & \\
\hline
	\end{tabular}} 
\end{table}
Those spectral levels are displayed also in Fig.~\ref{N11-Be11}, 
from which it is
evident that, in the low energy excitation range, the calculated excitation
energies agree well overall with observation. Table~\ref{Etable} also shows 
that
most calculated level widths compare with observations usually to within a 
factor of 2. 

Of note is that
there are calculated $\frac{7}{2}^-$ states at $\sim 4$ MeV excitation
in both $^{11}$Be and $^{11}$N that have no empirical 
counterpart in the range of energies shown. 
While there are no known experimental states at all in $^{11}$N above 5.08 MeV
excitation, there is a possible $\frac{7}{2}^-$ state
in the spectrum of $^{11}$Be at 6.705 MeV excitation. 
$\frac{7}{2}^-$ states exist in
$^{11}$C and $^{11}$B with excitation energies of 6.48 and 6.74 MeV
respectively with the calculated MCAS values being $\sim 1$ MeV 
less as shown in Fig.~\ref{C11-B11}.

These calculations involve very strong coupling, as evidenced by the value 
of $\beta_2 = -0.83$. Varying that value 
has a drastic effect, with a small decrease resulting in a significant raising 
of the centroid energy of the $\frac{1}{2}^+$ 
resonance for it to no longer be the ground state of $^{11}$N. On the other 
hand a small increase makes the $\frac{1}{2}^+$ 
ground state centroid decrease in energy. At the same time the width of that 
ground state resonance also decreases and a value 
of 800~keV can be found, in agreement with that listed~\cite{Ke12}. But then, 
on weakening the central potential strength to 
recover the wanted centroid energy  of 1.49 MeV, the width increases  back 
to over 3 MeV.

Allowing the chosen states of the core nucleus, $^{10}$C, to be orthonormal
admixtures including two quadrupole phonon
components, also can give a $\frac{1}{2}^+$ ground state resonance for $^{11}$N 
with a width of $\sim 800$ keV. But then
the centroid energy lies between 0.8 and 1.0 MeV. Again, weakening the central 
potential strengths to place that centroid at 1.49 MeV
leads to an increase in those widths to between 3 and 4 MeV.

\begin{figure}
\centering\scalebox{0.50}{\includegraphics*{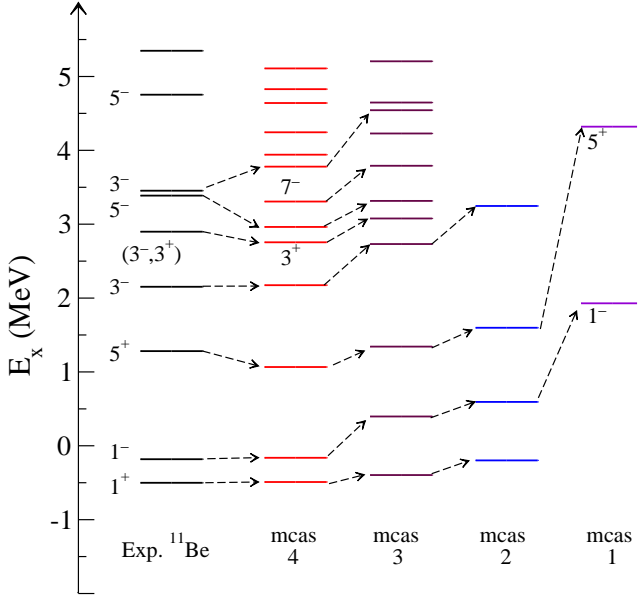}}
\caption{
\label{be11-sepst}
Experimental spectra of $^{11}$Be \cite{Ke12}
compared with the results of diverse calculations for
the $n+^{10}$Be cluster.
The spin-parities of the states are indicated
as $2J^\pi$.
The numerals under ``MCAS'' indicate reduction in the number of
target states considered in each calculation
as is indicated in the text.}
\end{figure}

In Fig.~\ref{be11-sepst}, we show how the strong coupling,
particularly to the $2_1^+$ state of the core, $^{10}$Be,
leads to the ground state in $^{11}$Be having a spin-parity
$\frac{1}{2}^+$. Therein the low excitation known spectrum of $^{11}$Be
is compared to the full coupled-channel result, as displayed previously in
Fig.~\ref{N11-Be11}, and here labelled as ``MCAS 4'' with
the numeral being the number of states of $^{10}$Be used in forming the 
coupled-channel interaction matrix
of potentials, to wit the ground $0^+$, and the $2_1^+, 2_2^+$ and $0_2^+$ 
in sequence.
The adjacent columns show the spectra formed with
just the first three of those (``MCAS 3''), 
with only the ground and $2_1^+$ (``MCAS 2''),
and as a single channel, ground state, calculation (``MCAS 1'').
Crucially all other details involved in these calculations
were unchanged.
Clearly the major spectral change with the evaluated
ground state being a $\frac{1}{2}^+$ level
arises from the coupling involving the $2_1^+$
state but the other terms play important roles
in adjusting evaluated energies downward with
improvement of the energy gaps in the
low lying $^{11}$Be spectrum.
Notably, the shell model calculations presented in Section~\ref{mass11iso}, 
Table~\ref{specamp}, 
also indicate this large contribution of the $2_1^+$ state.


\subsection{The mirror systems, $^{11}$C and $^{11}$B, with $^{10}$B as core}

Treating these nuclei as the clusters of a proton and 
of a neutron on the core nucleus, $^{10}$B, is a
major problem.   First the low energy spectrum of $^{10}$B
does not readily map to any simple collective model of
structure with its ground state spin-parity of $3^+$
being a challenge to be found even in larger space 
shell-model studies (see below).  

The  six lowest lying states in $^{10}$B have the 
spin-parity sequence $3^+_1, 1^+_1, 0^+, 1^+_2, 2^+$ and 
$3_2^+$; a sequence which is not easily identifiable with
collective structure.  Thus, a simple rotation scheme with 
solely a quadrupole coupling between all states was 
used to see what might result from our coupled-channel 
approach. The Woods-Saxon potential and 3pF Coulomb charge distribution
parameter values used are shown in Table~\ref{Param3}.
\begin{table}
\caption{\label{Param3} 
The parameter values used in the coupled-channel
evaluations of the $^{11}$C (as $p+^{10}$B) and $^{11}$B (as $n+^{10}$B)
spectra.
Strengths are in MeV, radii and diffusivities are in Fermi.}
\begin{tabular}{cccc}
\hline
	\rule[-0.2cm]{0pt}{0pt} \rule{0pt}{0.4cm}
\rule{0pt}{0.4cm}
Nuclear (WS)  &  $R_{0} = 2.8$   & $a_{0} = 0.65$ & $V_0 = -41.7$\\ 
	& & & $V_{ls} = \phantom{-}6.2$ \\
	& & & $\beta_2 = -$0.8\\
\rule{0pt}{0.4cm}
\rule[-0.2cm]{0pt}{0pt}
Coulomb (3pF)  & $R_{c} = 2.355$ & $a_{c} = 0.522$ & $w_c = -0.15$ \\
\hline
\end{tabular}
\end{table}

\begin{figure}
\scalebox{0.55}{\includegraphics*{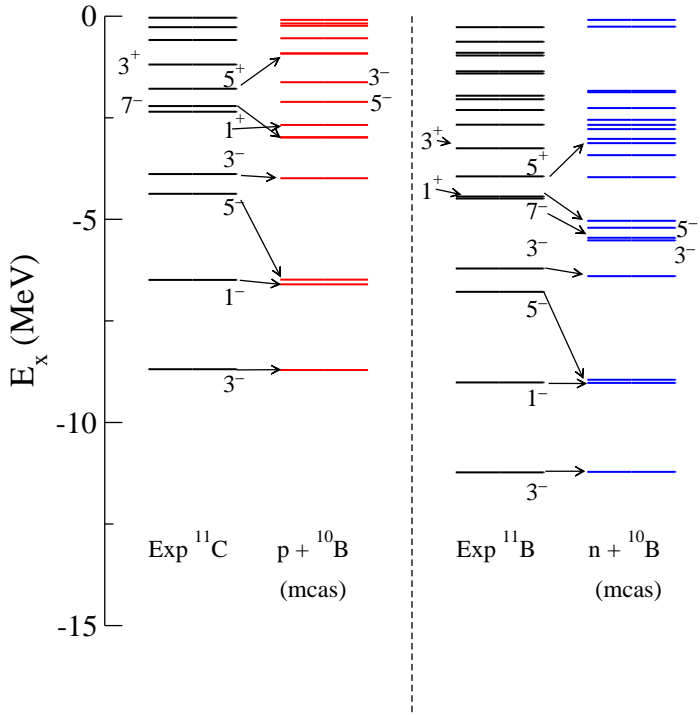}}
\caption{
\label{B10+np}
Experimental spectra of $^{11}$C and of $^{11}$B 
compared with the results of MCAS calculations for
the $p+^{10}$B and $n+^{10}$B clusters.  The zero 
energy in each is the relevant nucleon separation 
energy.  The spin-parities of the states are indicated
as $2J^\pi$.} 
\end{figure}

The OPP strengths needed to 
achieve the final results are listed in Table~\ref{OPP3}.
\begin{table}
\caption{\label{OPP3}
OPP strengths for  nucleon interactions with states in 
$^{10}$B required to get the results shown in 
Fig.~\ref{B10+np}.  All energies are in MeV.}
	\scalebox{0.95}{%
	\begin{tabular}{ccccc}
\hline
\rule[-0.2cm]{0pt}{0pt} \rule{0pt}{0.4cm}
state & \hspace*{0.2cm} $E_x$, $^{10}$B \hspace*{0.2cm}  
& \hspace*{0.4cm} $0s_{1/2}$ \hspace*{0.2cm} & 
\hspace*{0.2cm}$0p_{3/2}$ \hspace*{0.2cm} &
\hspace*{0.2cm} $0p_{1/2}$ \hspace*{0.4cm} \\
\hline
\rule{0pt}{0.4cm}
g.s. $3^+$   & 0.000  & $10^6$ & 20.0 & 10.0 \\
$1_1^+$      & 0.718  & $10^6$ & 4.0  & 10.0  \\
$0_1^+$      & 1.740  & 13.5   & 4.0  & 9.0  \\
$1_2^+$      & 2.154  & $10^6$ & 0.0  & 0.0 \\
$2_1^+$      & 3.587  & $10^6$ & 0.0  & 0.0 \\
\rule[-0.2cm]{0pt}{0pt} 
$3_2^+$      & 4.774  & $10^6$ & 0.0  & 14.0 \\
\hline
	\end{tabular}}
\end{table}
These values reflect underlying nucleon shell occupancies 
disparate to those in the core nuclei of the other clusters 
considered. Most notably we need to have some inner core 
($0s$-shell) breaking with the monopole.  This breaking is 
essential to get the known $\frac{1}{2}^+$ in the calculated 
result. Comparisons with the known spectra~\cite{Ke12}
(of $^{11}$C and $^{11}$B) are made in Fig.~\ref{B10+np}.
The low excitation spectra are reasonably in agreement,
with only the calculated $\frac{5}{2}^-$ state in the 
calculated spectrum of $^{11}$C being over an MeV too low and too close 
to the first excited $\frac{1}{2}^-$ state.

\section{Shell-model study of mass-10 and mass-11 isobars}
\label{sect3}

As a comparison to the spectra for the mass-11 pairs obtained 
from the coupled-channel evaluations, alternate spectra were 
obtained using the shell model. For the mass-10 nuclei, a complete 
$(0+2+4)\hbar\omega$
shell-model calculation was performed using the Zheng $G$-matrix 
interaction \cite{Zh95}. 
For the mass-11 systems, the Millener-Kurath
interaction \cite{Wa89} was used in a complete $(0+2)\hbar\omega$ model space. 
The positive parity states of $^{11}$Be and $^{11}$N were obtained in a 
$(1+3)\hbar\omega$
model space. The OXBASH shell-model program \cite{Ox86} was used 
to obtain both the wave functions and spectra.

\subsection{Mass 10 isobars}

\begin{figure}
\centering\scalebox{0.4}{\includegraphics*{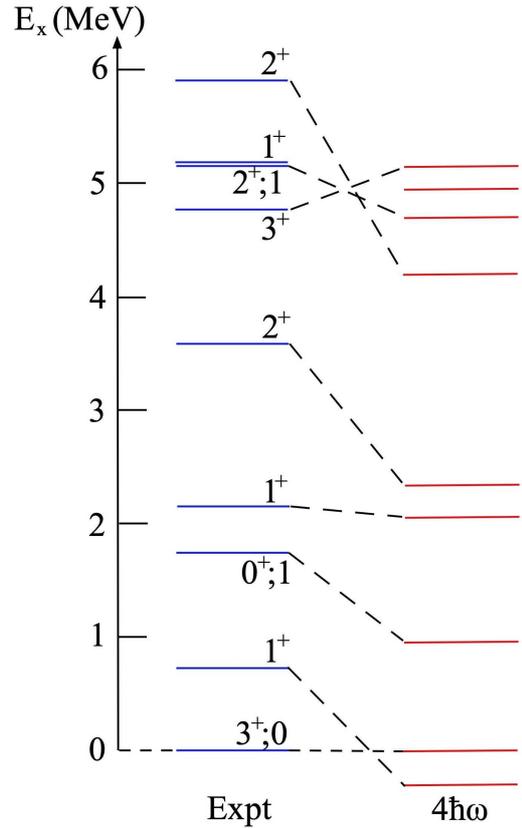}}
\caption{\label{b10_spec} Low-energy spectrum of $^{10}$B. 
The experimental spectrum \cite{Ti04} is compared to the results 
of the shell-model calculation as described in the text.}
\end{figure}

\begin{figure}
\centering\scalebox{0.45}{\includegraphics*{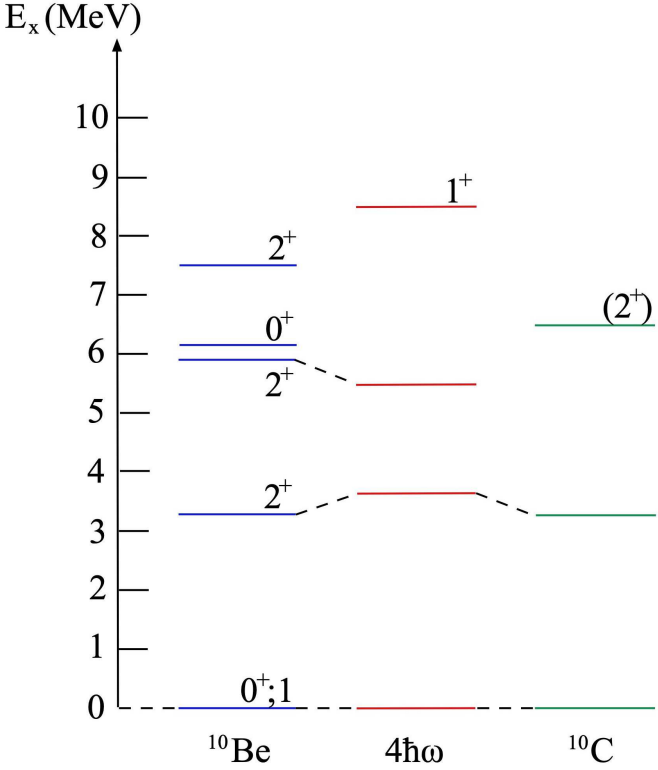}}
\caption{\label{be10_c10_spec} Low-energy spectra of the mirror 
pair $^{10}$Be and $^{10}$C. The data \cite{Ti04} are compared 
to the results of the shell-model calculation as described in 
the text.}
\end{figure}

In Figs.~\ref{b10_spec} and \ref{be10_c10_spec} the 
low-energy spectra for $^{10}$B and the mirror pair 
$^{10}$Be-$^{10}$C are shown, respectively. Therein, the results 
of the shell-model calculations are compared to the known spectra~\cite{Ti04}.

The nucleus $^{10}$B is a special case, being one of the few 
light odd-odd mass nuclei. Its ground state has a spin-parity of
$3^+$ which presents a problem for shell-model descriptions. 
 It is not a simple matter of the coupling of the odd
$0p_{3/2}$ proton to the odd $0p_{3/2}$ neutron. The result 
from the $(0+2+4)\hbar\omega$ shell model gives a 
ground state of $1^+$, but with a binding energy only 319~keV below 
that of the predicted $3^+_1$ state. The 
wave functions of those states are, as given by the
shell-model calculation:
\begin{align}
\left| 1^+_1 \right\rangle & = 69.21\% \left| 0\hbar\omega \right\rangle + 
17.18\% \left| 2\hbar\omega \right\rangle
+ 13.61\% \left| 4\hbar\omega \right\rangle \nonumber \\
\left| 3^+_1 \right\rangle 
& = 67.79\% \left| 0\hbar\omega \right\rangle + 17.75\% 
	\left| 2\hbar\omega \right\rangle + 14.28\%
\left| 4\hbar\omega \right\rangle \, ,
\label{comps}
\end{align}
while the dominant $0\hbar\omega$ configurations are
\begin{align}
\left| 1^+_1 \right\rangle & = 28.63\% \left| \left( 0p_{3/2} \right)^5 
\left( 0p_{1/2} \right) \right\rangle \nonumber \\
& \ \ \ \ + 28.61\% \left| \left( 0p_{3/2} \right)^4 
	\left( 0p_{1/2} \right)^2 \right\rangle  \nonumber \\
& \ \ \ \  + 7.87\% \left| \left( 0p_{3/2} \right)^3 \left( 0p_{1/2} 
	\right)^3\right\rangle 
+ 3.21\% \left| \left( 0p_{3/2} \right)^6 \right\rangle \nonumber \\
\left| 3^+_1 \right\rangle & = 26.00\% \left| \left( 0p_{3/2} \right)^6  
	\right\rangle 
+ 21.26\% \left| \left( 0p_{3/2} \right)^5 \left( 0p_{1/2} \right) 
	\right\rangle \nonumber \\
& \ \ \ \  + 18.42\% \left| \left( 0p_{3/2} \right)^4 \left( 0p_{1/2}  
\right)^2 \right\rangle\, .
\label{compsa}
\end{align}
A closed $0s_{1/2}$ shell forms part of each component listed in 
Eqs.~(\ref{compsa}). Other configurations not
explicitly given in Eqs.~(\ref{compsa}) are $\sim 1\%$ or less, leading 
to both wave 
functions being $\sim 70$\%
$0\hbar\omega$ with the rest of the wave functions formed from higher order
$\hbar\omega$ components.
 
There is significant occupation of the $0p_{1/2}$ orbit in both wave 
functions, while a dominant $(0p_{3/2})^6$ component is only in the 
wave function for the $3^+_1$ state. While a $1^+$ ground state is favored
in such a model, as is suggested by the dominant configurations involving
the $0p_{1/2}$ orbit in each wave function, the small excitation energy
of the  $3^+$ state encourages the idea that 
a small change in the 
shell-model interaction matrix elements could invert the
levels for the $3^+$  to be the ground state.  
This problem has also been considered using other  
shell models, such as by those classified as {\it ab initio}~\cite{Ca02,Ch20}.
Both of these studies consider very large basis spaces but do not
 account for all possible states contained therein. Thus the spaces are
incomplete and so require adjusting methods, such as that of the 
projection method of Gloeckner and Lawson~\cite{Gl74}, to account
for spurious center of mass motion. Nonetheless, both studies obtained
good spectra and electromagnetic properties of the systems studied, for
the mass-10 isobars~\cite{Ca02} and for the isotopes of Boron~\cite{Ch20}.
The first of these \cite{Ca02} used the CD-Bonn and Argonne realistic 
$NN$ interactions, but in the case of $^{10}$B, even allowing 
changes in the oscillator 
energy defining the single particle states, still gave a $1^+$ 
ground state.
They considered that as an  indication of
the need for true three-body forces to describe the low-lying
structure in complex nuclei.
 That was considered in the more recent study~\cite{Ch20} 
 and found to be the case but  reproduced using one of the
set of starting $NN$ interactions, the INOY NN interaction.

In our $(0+2+4)\hbar\omega$ calculation,
the dominant component of the $0^+_1$ state in $^{10}$B is 
$70\% \left| \left( 0s_{1/2} \right)^4 \left( 0p_{3/2} 
\right)^6 \right\rangle$, suggesting a closed $0s$ core. However, the 
occupation number is 1.9 for both the protons and neutrons in the
$0s$ shell; an effect of 
the other 30\% of the wave function 
which contains either an open $0s_{1/2}$ shell and/or occupancy in 
the $0d1s$ shell. 
Such result provides a microscopic
justification of a small  OPP strength parameter for 
for the $0^+_1$ level of the $^{10}$B system, as reported in
Table~\ref{OPP3}.

The low-energy spectra for the mirror nuclei $^{10}$Be and 
$^{10}$C are shown in Fig.~\ref{be10_c10_spec}, with the result 
of the calculation using the $(0+2+4)\hbar\omega$ shell model. 
The spectra are reasonably well reproduced, reflecting the mirror
symmetry, although there is not much experimental information 
about the spectrum of $^{10}$C. The first excited $0^+$ state in 
$^{10}$Be is predicted to be at 10.779~MeV, well above the 
measured energy. The wave function of this particular 
$0^+$ state is
\begin{equation}
\left| 0^+_2 \right\rangle = 71.42\% \left| 0\hbar\omega 
\right\rangle 
+ 15.15\% \left| 2\hbar\omega \right\rangle
+ 13.43\% \left| 4\hbar\omega \right\rangle.
\end{equation}
By comparison, the energy of that state is predicted to be 
between 9.51~MeV and 9.78~MeV in the model of Caurier 
\textit{et al.} \cite{Ca02}, and is also largely a 
$0\hbar\omega$ state.  However, this $0^+$ state, at 6.179~MeV, 
is considered to be predominantly $2\hbar\omega$ in 
nature \cite{Mi01}.

\subsection{Mass 11 isobars}
\label{mass11iso}

The spectra of the mass-11 isobars of interest, the mirror pairs 
$^{11}$B/$^{11}$C and $^{11}$Be/$^{11}$N, have been obtained from 
MCAS evaluations as shown in Figs.~\ref{C11-B11}, \ref{N11-Be11}, 
and \ref{B10+np}. The mirror pair $^{11}$Li/$^{11}$O presents a 
different problem for the shell model, given the very loose binding 
of $^{11}$Li (the two neutron separation energy is 369.1~keV while 
the single neutron separation energy is 395.5~keV \cite{Ke12}), and 
$^{11}$O is unbound.  The spectra of these nuclei of interest have 
also been obtained using a $(0+2)\hbar\omega$ shell model. In the
instances where odd-parity states are required, the state have been
obtained in a $(1+3)\hbar\omega$ model. 
The MK3W interaction of Warburton and 
Millener \cite{Wa89}, a modification of the earlier shell-model 
interaction of Millener and Kurath, was used.

\begin{figure}
\centering\scalebox{0.45}{\includegraphics*{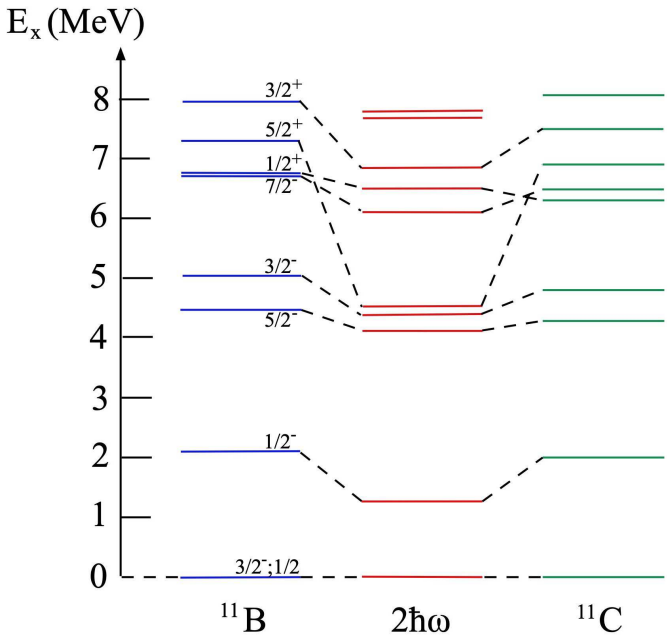}}
\caption{\label{b11c11_spec} Low-energy spectra of the mirror system 
$^{11}$B/$^{11}$C. The results of the $(0+2)\hbar\omega$ shell-model 
calculation are compared to the data \cite{Ke12} for both nuclei.}
\end{figure}

The low-energy spectra for the mirror pair $^{11}$B/$^{11}$C is 
shown in Fig.~\ref{b11c11_spec}.
Therein, the results of the $(0+2)\hbar\omega$ shell model calculation 
agree reasonably well with the measured spectra, with the exception 
of the $\frac{5}{2}^+$ state at 7.29~MeV in $^{11}$B and 6.91~MeV in 
$^{11}$C. The prevalence of negative parity states in this region of 
excitation of both these nuclei would indicate the need for larger 
shell-model spaces and the inclusion of higher $\hbar\omega$ 
components.

\begin{figure}
\centering\scalebox{0.45}{\includegraphics*{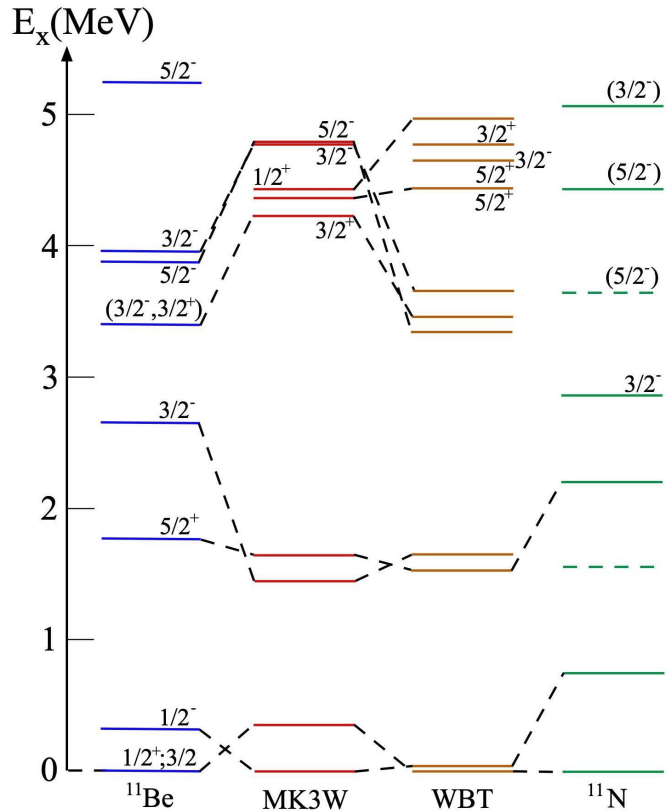}}
\caption{\label{be11n11_spec} Low-energy spectra for the mirror pair 
$^{11}$Be/$^{11}$N. The data \cite{Ke12} are compared to the results
of the shell-model calculations made in a $(0+2)\hbar\omega$ model 
space using the MK3W \cite{Wa89} and WBT \cite{Wa92} shell-model 
interactions.}
\end{figure}

Fig.~\ref{be11n11_spec} displays the low-energy spectra for the mirror 
pair $^{11}$Be/$^{11}$N. 
Not much is known of the spectrum of $^{11}$N, 
but the spectrum of $^{11}$Be is well-established as a result of 
measurements of the $\beta$-decay of $^{11}$Li \cite{Ke12}.
The presence of the $\frac{1}{2}^+$ ground state in both $^{11}$Be and 
$^{11}$N presents a problem for the shell model.  The result of the 
$(0+2)\hbar\omega$ shell-model calculation using the MK3W interaction 
predicts a $\frac{1}{2}^-$ ground state, with the $\frac{1}{2}^+$ state 
320~keV above it. For comparison, we have also performed a calculation 
in the same model space using the WBT interaction of Warburton and 
Brown \cite{Wa92}. It gives the $\frac{1}{2}^+$ ground state 
with the $\frac{1}{2}^-$ state only 20~keV in excitation. The ground state
configuration in the WBT model is
\begin{align}
\left| \frac{1}{2}^+_{\text{gs}} \right\rangle & = 
38.54\% \left| \left( 0s_{1/2} \right)^4 \left( 0p_{3/2} \right)^6 
	\left( 1s_{1/2} \right) \right\rangle \nonumber \\
& + 26.44\% \left| \left( 0s_{1/2} \right)^4 \left( 0p_{3/2} \right)^4 
	\left( 0p_{1/2} \right)^2 \left(
1s_{1/2} \right) \right\rangle \nonumber \\
& + 5.78\% \left| \left( 0s_{1/2} \right)^4 \left( 0p_{3/2} \right)^5 
	\left( 0p_{1/2} \right) \left( 1s_{1/2} \right)
\right\rangle \nonumber \\
& + 3.09\% \left| \left( 0s_{1/2} \right)^4 \left( 0p_{3/2} \right)^2 
	\left( 0p_{1/2} \right)^4 \left( 1s_{1/2} \right) \right\rangle
\nonumber \\
& + 26.15\% \text{ configurations not involving the } 1s \text{ orbit,}
\label{be11gs}
\end{align}
with the breakdown in $\hbar\omega$ being
\begin{equation}
\left| \frac{1}{2}^+_{\text{gs}} \right\rangle = 94.29\% \left| 
	1\hbar\omega \right\rangle 
+ 5.71\% \left| 3\hbar\omega \right\rangle.
\end{equation}
The configurations involving a nucleon in the $1s$ orbit account for 
73.85\% of the total wave function
within which the occupation numbers for the 
$1s_{1/2}$ orbit are 0.787 for neutrons and 0.006 for protons.
These components may be viewed as forming a $1s$ 
neutron halo in $^{11}$Be, but there is a substantial component
(26.15$\%$) that has no $1s_{1/2}$ involvement. Strong configuration mixing
is suggested by Eq.~(\ref{be11gs}); a result enhanced
with the shell model calculations made
using the MK3W interaction from which
configurations involving the $1s_{1/2}$ orbit
account for only 49.48\% of the total wave function.

The relevant couplings of the valence neutron to states in $^{10}$Be
forming the ground state in $^{11}$Be may also be obtained from the shell model
calculation. The spectroscopic amplitudes for forming the ground
state of $^{11}$Be as obtained from the shell model calculations
using the WBT and MK3W interactions are given in Table~\ref{specamp}.
\begin{table}
\caption{\label{specamp} Shell-model spectroscopic amplitudes, $S_j$, for 
	the coupling of a
neutron to states in $^{10}$Be forming the $\frac{1}{2}^+$ ground state 
	in $^{11}$Be.}
	\scalebox{1.0}{%
\begin{tabular}{c|ccc}
	\hline
	\hspace*{0.4cm} State \hspace*{0.4cm} 
	& \hspace{0.4cm} Orbit \hspace{0.4cm} &
	Interaction & $S_j$ \\
\hline
g.s $(0^+)$ & $1s_{1/2}$ & WBT & $-$1.2293 \\
		    &     & MK3W & \phantom{-}1.1097 \\
	$2^+_1$ & $0d_{5/2}$ & WBT & \phantom{-}0.5538 \\
		 &        & MK3W & \phantom{-}0.6317 \\
 $2^+_1$                 & $0d_{3/2}$ & MK3W & $-$0.1122 \\
\hline
\end{tabular}}
\end{table}
The results of the shell model calculations show strong coupling of the 
neutron not only to the ground
state of $^{10}$Be but also to the $2^+_1$ state. 
In the case of the WBT shell model calculation 
the coupling to the $2^+$ state is almost purely from the neutron occupying 
the $0d_{5/2}$
orbit. The MK3W result, however, indicates an additional coupling via the 
neutron occupying the $0d_{3/2}$
orbit. The spectroscopic amplitude in that case is of opposite sign, 
indicating an overall weaker
coupling to the $2^+$ state in the model. This indication of a strong 
coupling to the ground and
$2^+$ states supports the necessity of such coupling in the MCAS calculation
(cf. Table~\ref{Param2}).

As with the ground state of $^{10}$B, a small change in the matrix elements
in both the MK3W and WBT interactions may give the 
correct inversion with the correct energy separation.  Note that the states in 
$^{11}$Be were not used in the 
determination of the WBT interaction \cite{Wa92}. 

The spectrum obtained using the WBT interaction also confirms the results 
obtained in
Ref.~\cite{Sa93}. Therein, the authors note that it is possible for a shell 
model to
reproduce the parity inversion in $^{11}$Be, but also ascribe the inversion 
to the coupling
of the valence neutron to the $2^+$ state in the $^{10}$Be core. This supports 
the inclusion
of that state as part of the target spectrum used
in our coupled-channel calculations of $^{11}$Be and $^{11}$N. 
Though less is known
about the resonance states  in $^{11}$N, 
while being indicative of some mirror 
symmetry with $^{11}$Be,   
the results of the 
shell-model calculations show tentative agreement with some of the 
known spin-parities of those resonances in the spectrum.

The parity inversion (of the ground and first excited states) observed 
in both $^{11}$Be and $^{11}$N has also been the subject of other shell 
model studies
\cite{Sa93,Ca16};  the latter of which noted that the effects of quadrupole 
core excitation and pairing are important to find the parity inversion in 
$^{11}$Be.
The former however performed essentially a traditional microscopic calculation 
but over many shells, and  
with their optimised interaction for the chosen model space, found the desired 
inversion.
 

\section{Conclusions}
\label{conc}

The low-energy spectra of four mass-11 nuclei ($^{11}$Be, $^{11}$B, $^{11}$C, 
and $^{11}$N) have been evaluated,
initially, by using the multi-channel algebraic scattering approach treating 
the nuclei as the coupling of a nucleon
with states of a mass-10 nuclear core. Coupled-channel Hamiltonians were 
formed assuming a collective model
of the nuclear cores and of their interactions with the extra-core nucleon. 
A vibration model was used to specify the
target states of $^{10}$Be and $^{10}$C
as their low excitation
spectra have the essential sequence of states for that description.
Mirror symmetries were assumed and the Coulomb potentials for interactions 
with protons
were taken from three-parameter Fermi descriptions of the nuclear charge 
distributions.
In contrast, with $^{10}$B as the core, a rotation model was used 
to specify the interaction potential matrices input to MCAS calculations
since the low--excitation spectra of $^{10}$B does not show
any semblance of a vibration model structure. However the spectra of $^{11}$B 
and $^{11}$C, described 
by the rotational model with the odd-odd $^{10}$B core, set the second 
excited state, the $\frac{5}{2}^-$ level, 
too close to the first excited state, $\frac{1}{2}^-$, making them to appear 
almost degenerate. 
This characteristic is not observed in the measured spectra of both mirror nuclei.

The vibration couplings to the collective motion of the  even-even core states 
were important especially in the description of
the states in $^{11}$Be and $^{11}$N. Those nuclei have positive-parity ground 
states, opposite to those
of the ground states of the neighbouring mass-11 nuclei, with  $^{11}$Be having 
a single-neutron halo ground state. From phenomenological 
descriptions of that nucleus, significant coupling to the $2^+$ state in 
$^{10}$Be is required in order to
obtain the $\frac{1}{2}^+$ ground state. Such coupling is naturally included 
in the coupled-channel approach used.

The low-energy spectra of both the mass-10 and mass-11 
nuclei considered were also described using complete  basis space
shell models. Specifically no-core $(0 + 2)\hbar\omega$ and 
$(0 + 2 + 4)\hbar\omega$ space shell-model studies were made.
The complete $(0 + 2 + 4)\hbar\omega$ space 
was used to define the spectra of the mass-10 systems
with the Zheng $G$-matrix interaction. For the
mass-11 systems, the fitted interactions (MK3W and WBT) 
were used in the complete $(0 + 2)\hbar\omega$ model space. 
The low-energy spectrum for
$^{11}$Be obtained from the shell model calculation using the WBT interaction 
gave
the correct spin-parity of the ground state, but not the energy of the 
first excited state.
However, it is important to note that in that model significant occupation 
of a single
neutron in the $1s_{1/2}$ orbit is found, consistent with the single-neutron 
halo description of
the ground state of $^{11}$Be. Also, the spectroscopic amplitudes for the 
coupling
of a neutron to states in $^{10}$Be as obtained by the shell model support 
the significant coupling of
the neutron to both the ground and $2^+$ states in $^{10}$Be as required by MCAS.

The results of our calculations of the mirror systems, $^{11}$C and $^{11}$B
treated as the clusters $n+^{10}$C and $p+^{10}$Be respectively,
gave excellent agreement with the known spectra of these nuclei to over
10~MeV excitation and with the known nucleon separation
energy. For $^{11}$N and $^{11}$Be treated as the clusters
$p+^{10}$C and $n+^{10}$Be, however,  the calculated results are not perfectly
matched to the known spectra. But the sequencing and, notably,
nucleon emission thresholds, are given correctly.  Unlike
$^{11}$C and $^{11}$B, $^{11}$Be is weakly 
bound, while  $^{11}$N is unbound.
These results, involving clusters coupling to form states in the 
weakly bound $^{11}$Be and unbound $^{11}$N, may indicate inadequacy 
in the simple collective model used to specify the coupled-channels 
Hamiltonian.  The nuclei, $^{11}$C and $^{11}$B, considered as the 
clusters $p+^{10}$B and $n+^{10}$B respectively, were found to give
reasonable spectra again to over 10~MeV excitation, notwithstanding
a breaking 
of the $0s_{1/2}$ shell in the specification of the OPP of the excited 
$0_1^+$ state of the core which was found
necessary to have the correct location of the 
$\frac{1}{2}^+$ state. That requirement is, in part,
supported by the results of the shell-model calculation for $^{10}$B.

Aside from there being insufficient data to provide
constraints on the details of the structure model for $^{10}$C,
with even the spin-parities of two of those four states selected
in the coupled-channel calculations being not listed in the
data sheets, the assumed vibration model
of structure for the core nuclei may be too simplistic.
Improved structure models for $^{10}$Be and $^{10}$C need to be
used before a higher quality agreement between known spectra
with these cluster model results may be achieved.

\section*{Acknowledgments}
We thank the Staff at the High-Performance Computing Centre of the University
of Melbourne, in particular Lev Lafayette, Sean Crosby, Greg Sauter, Linh Vu
and Bernard Meade, for the use of SPARTAN~\cite{La16} in making all the
calculations to find the MCAS results presented herein.


\end{document}